\documentclass[preprint,tightenlines,aps,floats,nofootinbib,amssymb,groupedaddress]{revtex4-1}
\usepackage[sort&compress]{natbib}
\usepackage{epsf,epsfig}
\usepackage{amsmath}
\usepackage{hyperref}
\usepackage{subfigure}
\usepackage{graphicx}
\usepackage{color}
\usepackage{mathrsfs}
\usepackage{booktabs}
\usepackage{multirow}

\newcommand{\be}{\begin{equation}}
\newcommand{\ee}{\end{equation}}
\newcommand{\ba}{\begin{array}}
\newcommand{\ea}{\end{array}}
\newcommand{\bea}{\begin{eqnarray}}
\newcommand{\eea}{\end{eqnarray}}

\newcommand{\bi}{\begin{itemize}}
\newcommand{\ei}{\end{itemize}}
\newcommand{\bal}{\begin{aligned}}
\newcommand{\eal}{\end{aligned}}

\makeatletter

\begin{document}
\title{Glueballs and deconfinement temperature in AdS/QCD}	
\author{S.S. Afonin}
\email{s.afonin@spbu.ru}
\author{A.D. Katanaeva}
\email{alice.katanaeva@gmail.com}
\affiliation{Saint Petersburg State University,\\
7/9 Universitetskaya nab., St.Petersburg, 199034, Russia}
\thispagestyle{empty}

\begin{abstract}
We put forward an alternative way to estimate the deconfinement temperature in bottom-up holographic models.
The deconfinement in anti-de Sitter (AdS)/QCD is related to the Hawking--Page phase transition, in which
the critical Hawking temperature is identified with the deconfinement one in the gauge theory.
The numeric estimation of the latter is only possible when the parameters of a five-dimensional dual model are
previously determined from consistency with other physical aspects, standardly, providing a description of
the QCD resonances. The traditional way to fix parameters in the simplest AdS/QCD models is to reproduce the mass of 
the $\rho$ meson or the slope of the approximate radial Regge trajectory of the $\rho$ excitations.
Motivated by a general idea that the slope value originates in gluodynamics,
we propose calculating the deconfinement temperature using the trajectory of scalar glueballs.
We consider several holographic models and use the recent idea of isospectral potentials to make
an additional check of the relevance of our approach. It is demonstrated that different models from an
isospectral family ({\it i.e.} the models leading to identical predictions for spectrum of hadrons with fixed
quantum numbers) result in different predictions for the deconfinement temperature. This difference is found to
be quite small in the scalar glueball channel but very large in the vector meson channel. 
The observed stability in the former case clearly
favours the choice of the glueball channel for thermodynamic predictions in AdS/QCD models. 
For a balanced approach, we argue that either assuming $f_0(1500)$ to have a dominating component of
$0^{++}$ glueball or accepting the idea of the universality in the radial Regge trajectories of light non-strange
vector mesons one can reproduce the results for the deconfinement temperature obtained before in the lattice simulations
in the background of non-dynamical quarks.
\end{abstract}

\maketitle

\section{Introduction}
The AdS/QCD correspondence proved to be a useful technique to study different facets of the 
strongly coupled theories.
Besides the established QCD content of mesons and baryons, one may try to apply the holography
to the hypothesized QCD bound states -- the glueballs. 

Introduced in 1970s \cite{Fritzsch:1972jv}, \cite{Fritzsch:1975tx}
these particles continue to be a subject of intense study both in lattice QCD and in various phenomenological models.
Experimental searches are carried out (BES III and Belle II) but no unambiguous identification may be claimed (see, for instance, the review 'non-$q\bar q$ Mesons' in Particle Data \cite{PDG} or \cite{Ochs:2013}).  
Future experiments, such as PANDA at FAIR facility and several programmes at NICA,
are expected to specify the existing hypotheses (see, {\it e.g.}, suggestions in \cite{Parganlija:PANDA} and \cite{Parganlija:NICA}).

Indeed, the holographic investigation of glueballs has been performed before: 
from first applications of the Maldacena conjecture ({\it e.g.} in \cite{Brower:2000rp}) and top-down realizations (\cite{Hashimoto:2007,Brunner:2015}, {\it etc.}), 
to bottom-up phenomenological approaches implemented in \cite{BoschiFilho:2002ta,COLANGELO200773,Forkel:2007,Colangelo:2007},
and more recent ones, like \cite{BoschiFilho:2012xr}, introducing more complicated models.
The focus of these papers mostly stays on predictions of the spectrum and decay rates, as well as on the corroboration
of the holographic correlation functions in the light of other existing QCD techniques.
In this work we would like to sidestep and provide the connection of this kind of bottom-up models to another pressing issue of holographic QCD.

The QCD phase diagram, another issue of interest in strong interactions, is subject to holographic treatment as well.
Within the bottom-up approach this type of studies was pioneered by C.P.~Herzog in \cite{Herzog:2006} and continued by
many authors (see, {\it e.g.}, references in \cite{estim_deconf}).
In these papers the $5D$ model is constructed from the point of view that the gravity part of the $5D$ action
describes the gluodynamic background and thermodynamic properties of the $4D$ holographic dual. 
The Hawking--Page type phase transition between different AdS backgrounds may happen at some critical temperature $T_c$,
then this characteristic is interpreted as the temperature
of transition from the confined to the deconfined phase. Additionally, the $5D$ matter content represents the $4D$ states
bound by the strong interactions.

The two sectors of the $5D$ model turn out to be closely intertwined when concerning the value of the deconfinement temperature.
The estimation of the last depends on the model parameters which 'historically' were connected to the dominant
purpose of the first bottom-up AdS/QCD papers (\cite{HW_2005, daRold_2005, SW_2006}, {\it etc.})
-- to the description of light vector meson spectra. The estimates of $T_c$ in \cite{Herzog:2006}
follow the parameter values of these works.
Though this is a traditional choice there seems to be no particular reason why the vector meson spectra should 
in any way determine the deconfinement temperature. In the same time, particle spectra, especially the radial Regge 
trajectories appearing in Soft Wall (SW) models, are one of the most attractive features of holography. 

We would like to propose a different kind of particle -- a spin zero glueball (and its radial excitations) -- 
to fix the matter part of the holographic model and define the model parameters.
To vindicate the concept we provide several 'pro' arguments in the text, that may be shortly given as:
\begin{itemize}
 \item The phase diagram can be studied in pure gluodynamics.
 Since the holographic approach is defined in the large-$N_c$ (planar) limit of gauge theories
 where the glueballs dominate over the usual mesons and baryons (as the quarks are in the fundamental
 representation), the gluodynamics must dictate the overall mass scale and thereby the major contribution
 to the deconfinement temperature $T_c$. 
 \item Within the isospectrality concept \cite{Vega_Cabrera}, one can show that the predicted values of $T_c$ are more stable 
 for scalar glueballs than for vector mesons.
 \item Considering phenomenological reasons, numerical values of $T_c$ determined in the scalar glueball framework
 can be interpreted as better fitting the lattice expectations.
\end{itemize}

The first argument was scrutinized in our paper \cite{estim_deconf} and we will further substantiate the point. One can observe, for example, 
that if we take the linear radial spectrum of scalar glueballs given by the standard SW holographic model, 
$m_n^2=\mu^2(n+2)$, $n=0,1,2,\dots$  (the interpolating operator $G_{\mu\nu}^2$ is assumed \cite{COLANGELO200773}), and consider the scalar resonance
$f_0(1500)$ \cite{PDG} as the lightest glueball (as is often proposed in the hadron spectroscopy \cite{PDG}) we will obtain the slope $\mu^2=1.13$ GeV$^2$.
It agrees perfectly with the mean radial slope $\mu^2=1.14\pm 0.01$ GeV$^2$ found for the light mesons in the analysis \cite{bugg} and achieved independently
in \cite{Afonin:unflmeson, Afonin:classicsym}.
In Ref. \cite{estim_deconf} we showed that this value of $\mu^2$ leads to a consistent numerical estimation of the deconfinement temperature in 
the SW model.

The second argument is motivated by an interesting finding of the authors of \cite{Vega_Cabrera} that the particular form
 of spectrum in SW-like models does not fix a model itself: 
One can always construct a one-parametric family of SW models (controlling modifications of the 'wall') leading to the same spectrum. We will demonstrate that such isospectral models, however,
result in different predictions for the deconfinement temperature $T_c$. In the vector channel  originally considered by Herzog in \cite{Herzog:2006} these variations can
be quite significant, whereas in the scalar case the difference may be rather small and spanning
an interval admissible by the accuracy of the large $N_c$ limit.

The third argument is just a phenomenological observation for typical predictions of $T_c$ in the bottom-up holographic models. For instance, the original Herzog's 
analysis of the vector Hard Wall (HW) holographic model with the $\rho$-meson taken as the lowest state resulted in the prediction $T_c=122$ MeV \cite{Herzog:2006}.
If we apply this analysis to the scalar HW model with $f_0(1500)$ as the lightest glueball, we will find $T_c \approx 150$ MeV. 
The lattice simulations typically predict the lightest glueball near $1.7$ GeV ($1.6 - 1.7$ GeV as quoted in Particle Data reviews \cite{PDG}). This value shifts the prediction to $T_c \approx 170$ MeV.
So we may regard the interval $T_c = 150 - 170$ MeV as a prediction of the HW model in the glueball channel. 
We find remarkable that exactly this interval was found in the modern lattice simulations  with dynamical
quarks \cite{Borsanyi:2010bp}. 

In the present paper, we will consider scalar glueballs and vector mesons in parallel
in order to demonstrate in detail the emergent differences. 
In Section \ref{Htemp} we recall the general procedure to achieve an estimation of the deconfinement 
temperature in AdS/QCD models. Then we provide some particular calculations in various not overcomplicated $5D$ frameworks in Section~\ref{holo_models}.
We introduce the notion of isospectrality in Section~\ref{isospectrality} and show how it can affect
the temperature values.
In Section~\ref{experiment} we briefly review lattice and experimental results in the relevant topics,
mostly to provide benchmarks and input parameters for the models considered. Finally, in Section~\ref{holo_T} we give our numerical
estimations of $T_c$ distinguishing the glueball and vector meson channels and several different options within.
We conclude in Section~\ref{conclusions} with a discussion on the general validity of the presented approach.

\section{Hawking temperature in bottom-up models}\label{Htemp}
Consider a general $5D$ action 
that contains a universal gravitational part and a matter part to be specified further,
and has an AdS related metric $g_{MN}$ ($g=\det g_{MN}$):
\bea
S &=& \int d^4xdz\sqrt{-g} f^2(z)\left(\mathcal L_{gravity} + \mathcal L_{matter} \right)\\
\mathcal L_{gravity} &=& -\frac1{2k_g} \left(\mathcal{R}-2\Lambda\right)
\eea
Here $k_g$ is the coefficient proportional to the $5D$ Newton constant, $\mathcal{R}$ is the Ricci scalar
and $\Lambda$ -- the cosmological constant.
The choice of the dilaton background, $f(z)$, distinguishes possible holographic models.
They differ as well by the interval the $z$ coordinate spans. For now we assume $z\in [0,z_{max}]$,
though $z_{max}=\infty$ is possible and will be of the main interest in the present work.

The assessment of the critical temperature is related to the leading contribution in the large $N_c$ counting,
that is the $\mathcal L_{gravity}$ part scaling as $\frac1{2k_g} \sim N_c^2$. $\mathcal L_{matter}$ scales 
as $N_c$ and thus does not explicitly affect the confinement/deconfinement process.

It is found out that the deconfinement in AdS/QCD occurs as a Hawking -- Page phase transition~\cite{Herzog:2006}.
Let us recall how the order parameter of this transition is defined for a given theory.

First, we should evaluate the free action densities $V$, being the regularized $S_{grav}$, 
on different backgrounds corresponding to two phases. 
One assumes that the thermal AdS of radius $R$ is defined by the general AdS line element
\be
ds^2=\frac{R^2}{z^2}\left(dt^2-d\vec{x}^2-dz^2\right),
\ee
with the time direction restrained to the interval $[0, \beta]$. This background corresponds to the confined phase.
The metric of the Schwarzschild black hole in AdS describes the deconfined phase and is given by
\be
ds^2=\frac{R^2}{z^2}\left(h(z)dt^2-d\vec{x}^2-\frac{dz^2}{h(z)}\right),
\ee
where $h(z)=1-(z/z_h)^4$ and $z_h$ denotes the horizon of the black hole.

With the cosmological constant in $5D$ AdS being $\Lambda=-6/R^2$,
both these metrics are the solutions of the Einstein equations and provide the same value of the Ricci scalar $\mathcal R = -20/R^2$.
Hence, the free action densities differ only in the integration limits,
\begin{align}
V_{\text{Th}}(\epsilon)&=\frac{4R^3}{k_g} \int_0^\beta dt\int_\epsilon^{z_{max}}dzf^2(z) z^{-5},\\
V_{\text{BH}}(\epsilon)&= \frac{4R^3}{k_g}  \int_0^{\pi z_h}dt\int_\epsilon^{\min(z_{max},z_h)}dzf^2(z)z^{-5}.
\end{align}
The two geometries are compared at $z=\epsilon$ where the periodicity in the
time direction is locally the same, {\it i.e.} $\beta=\pi
z_h\sqrt{h(\epsilon)}$. Then, we may construct the order parameter for the phase
transition,
\begin{equation}
\label{deltaV}
\Delta V = \lim_{\epsilon\rightarrow0}\left(V_{\text{BH}}(\epsilon)-V_{\text{Th}}(\epsilon)\right).
\end{equation}
The thermal AdS is stable when $\Delta V>0$, otherwise the
black hole is stable. The condition $\Delta V=0$ defines the critical temperature $T_c$ at which the
transition between the two phases happens through the definition of the
Hawking temperature $T=1/(\pi z_h)$.

However, to provide a numerical estimation of $T_c$ the usage of Eqn.~(\ref{deltaV}) is not enough
as it yields $z_h$ as a function of model dependent parameters -- $z_{max}$ and/or those possibly
introduced in $f(z)$. We must appeal to the matter sector $\mathcal L_{matter}$ to give physical meaning to
these parameters and to connect $T_c$ to a particular type of a holographic model.

\section{Introducing matter content in various $5D$ models} \label{holo_models}

Following the stated route, we would like to compare the results for the deconfinement temperature
in different cases, whether we consider a matter content of vector mesons or scalar glueballs.
We provide them in parallel, browsing through the general features first.

The $\mathcal L_{matter}$ for an Abelian (for simplicity) vector field $A_M(x,z)$ and a scalar field $\varphi(x,z)$ 
are given by
\begin{align}
 \mathcal L_{v} =& -\frac1{4g_5^2}g^{MP}g^{NQ}\left(\partial_M A_N-\partial_N A_M\right)\left(\partial_P A_Q-\partial_Q A_P\right),\\
\mathcal L_{sc} =& \frac1{2k_s}\left(g^{MN}\partial_M\varphi\partial_N\varphi-M^2_5\varphi^2\right),
\end{align}
where $k_s$ and $g_5$ are the normalization constants. The $5D$ field $\varphi$ is dual to $G_{\mu\nu}G^{\mu\nu}$ -- the lowest dimension QCD
operator  that bears the scalar glueball quantum numbers $0^{++}$. The general methods of the AdS/CFT
conjecture prescribe $M^2_5R^2=0$ as the conformal dimension of the dual $4D$ operator is $4$.
The vector field $A_M$ is dual to the QCD conserved current $\bar q \gamma_\mu q$ of dimension $3$. 
AdS/CFT prescribes such vector $5D$ field to be massless as well.
The standard holographic gauge choice is $A_z=0$.

In the following if possible we are going to use the unified notation for both cases introducing the spin parameter
$J$, with $J=0$ corresponding to the scalar case and $J=1$ to the vector one. 
\footnote{We presume the method to be valid for any spin, under condition that the $5D$ action is of the form discussed in \cite{Brodsky:2014}.
Nevertheless, there exists no clear notion on the treatment of higher spin states in the literature. See different approach in \cite{SW_2006}, 
though it fails to provide the correct result for scalar fields and hence cannot be used by us.}

Let us call the general spin $5D$ field $\Phi(x,z)$ (we suppress the Lorentz indices). Then, the
 equations of motions (EOM) for the Fourier transformation of these arbitrary spin $5D$ fields read as 
 follows \cite{Brodsky:2014}
\be\label{jEOM}
\partial_z \frac{f^2(z)}{z^{3-2J}}\partial_z \Phi(q, z) -\frac{M^2_5R^2}{z^{5-2J}}f^2(z)\Phi(q, z)+ \frac{f^2(z)}{z^{3-2J}} q^2 \Phi(q,z)=0.
\ee
Further, we consider $M^2_5R^2=0$ as befits our models. In \cite{Brodsky:2014}, it is discussed that for higher spin fields the term coming with $M^2_5$
may turn out to be $z$ dependent and it is a condition of LFHQCD (Light Front Holographic QCD)
that it takes a particular constant value. We will not go into this discussion as
for our cases of vector and scalar Lagrangians Eqn.~(\ref{jEOM}) provides the conventional EOM with $M^2_5R^2$ defined from the holographic dictionary as mentioned above.

The standard procedure to find the spectrum is to solve EOM as the eigenvalue problem imposing
$q^2$ to take the discrete values $q^2=M^2(n)$ ($n=0,1,2...$ being a discrete parameter).
Then, the Kaluza--Klein (KK) decomposition is given as
\be
\Phi(q,z)=\sum\limits_{n=0}^\infty \phi_n(z)\phi^{(n)}(q).
\ee
$\phi^{(n)}$ represents the tower of $4D$ excitations.
Their $z$-profiles are determined by demanding specific boundary conditions in the $z$ direction.
Thus, further investigation requires a specification of the model.

\subsection{HW option}
Let us start with the simplest option -- the Hard Wall model \cite{HW_2005}, \cite{daRold_2005}.
This model is characterized by $f^2(z)=1$ and an explicit cut-off of the $z$ direction at some finite position $z_{max}$.
The $z$-dependent solution of Eqn.~(\ref{jEOM}) $\phi_n(z)$ should be subject to the Dirichlet boundary condition in the UV $\phi_n(0)=0$ and
the Neumann one on the IR cut-off $\partial_z\phi_n(z_{max})=0$.
The appropriate solution is
\be
\Phi(q, z) \sim (qz)^{2-J} J_{|2-J|}(qz),
\ee
where $J_\alpha$ is a Bessel function of the first type.
As the recurrence relation for Bessel functions states $\partial_z (z^\alpha J_\alpha) = z^\alpha J_{\alpha-1}$,
the IR boundary condition translates into $J_{\alpha-1}(M(n)z_{max})=0$.
The first zeros of $J_0$ and $J_1$ fix the values of the ground states as
\begin{align}
 m^{J=1}_{HW}=M_{J=1}(0)=\frac{2.405}{z_{max}},\\
 m^{J=0}_{HW}=M_{J=0}(0)=\frac{3.832}{z_{max}}.
\end{align}

Concerning the particular consequences for the order parameter of the Hawking--Page phase transition,
the result of Eqn.~(\ref{deltaV}) in the HW model is
\be
\Delta V_{HW}=0\ \Leftrightarrow\ z_{max}^4=2z_h^4,
\ee
as has been found in~\cite{Herzog:2006}.
Then, the deconfinement temperature is determined as
\begin{align}
T^{J=1}_{HW} &= \frac{2^{1/4}}{\pi z_{max}}=\frac{2^{1/4} m^{J=1}_{HW}}{2.405\pi} = 0.1574 m^{J=1}_{HW},\\
T^{J=0}_{HW} &= \frac{2^{1/4}}{\pi z_{max}}=\frac{2^{1/4} m^{J=0}_{HW}}{3.832\pi} = 0.0988 m^{J=0}_{HW}.
\end{align}
Evidently, these values have different numerical prefactors and depend on the mass of the first resonances,
which are not having the same physical origin. Hence, we cannot expect to get a universal estimation of $T_c$ in HW models;
for the numerics, see Section~\ref{holo_T}.

\subsection{(G)SW option}
The main achievement of the Soft Wall model \cite{SW_2006} was the reproduction of the linear  Regge  and radial trajectories
for mesons. It is natural to hypothesize that a scalar glueball and its radial excitation may lie on the 
linear trajectory as well.

The traditional SW model is characterized by an infinite IR cut-off $z_{max}=\infty$ and the conformality is broken by the introduction of
the dilaton profile $f^2(z)=e^{-\kappa^2z^2}$.
On the UV brane the Dirichlet condition is imposed and good convergence is required in the IR (to be suppressed by the dilaton exponent).
The normalizable discrete modes of Eqn.~(\ref{jEOM}) are
\be \label{SWeig}
\phi_n(z) = \mathcal N_n (\kappa z)^{2-J+|2-J|} L^{|2-J|}_n(\kappa^2z^2),
\ee
where $L_n^m$ are the generalized Laguerre polynomials, and $\mathcal N_n$ are the normalization factors of no importance to us. 
For the discrete parameter $n=0,1,2,...$, we obtain
the SW spectra,
\be
M_{J}^2(n)=4\kappa^2\left(n+1+\frac{|2-J|-J}2\right),
\ee
where one should choose $J=0$ or $1$, and $M_5^2R^2=0$ is assumed from the beginning.

A natural generalization toward a vector meson or glueball spectrum with an arbitrary intercept parameter $b$,
\be\label{GSW_sp}
M_{J}^2(n)=4\kappa^2\left(n+1+b+\frac{|2-J|-J}2\right),
\ee
may be achieved using the Generalized Soft Wall profile proposed in \cite{GSW} (see also \cite{GSW2}):
\be
f^2(z)=e^{-\kappa^2z^2}U^2(b,J-1;\kappa^2z^2).
\ee
The modification consists in the Tricomi hypergeometric function  $U$ that provides the necessary free parameter $b$
to the spectrum but does not change the SW asymptotes in UV and IR.  
As $U(0,J-1;x)=1$, GSW with $b=0$ reduces to the usual SW.

The estimation of $\Delta V$ is similar to the one performed in \cite{estim_deconf} (though there the $U$-function has
a fixed second parameter, adjusting to the vector spectrum) and results in
\be
\Delta V_{GSW}=\frac{ \pi R^3}{2k_gz_h^3}\left[U^2(b,J-1;0)
-4(\kappa z_h)^4\int\limits_{\kappa^2z^2_h}^{\infty}dte^{-t}t^{-3}U^2(b,J-1;t)\right].
\ee
Taking $b=0$ one easily recovers the SW model result
\be
\Delta V_{SW} = \frac{\pi R^3}{k_g z_h^3}\left[\frac12+e^{-\kappa^2z^2_h}(\kappa^2z^2_h-1)
-\kappa^4z^4_h\int\limits_{\kappa^2z^2_h}^{\infty}dte^{-t}t^{-1}\right].
\ee
Numerically, one can find the value of $z_h$, as a function of $\kappa$ (and $b$),
that solves the equation $\Delta V_{(G)SW}=0$.
For the simple SW we may reproduce the expression of the deconfinement temperature from \cite{Herzog:2006}
\be
T_{SW} \simeq 0.49\cdot\kappa.
\ee

\begin{figure}[t]
	\includegraphics[scale=0.45]{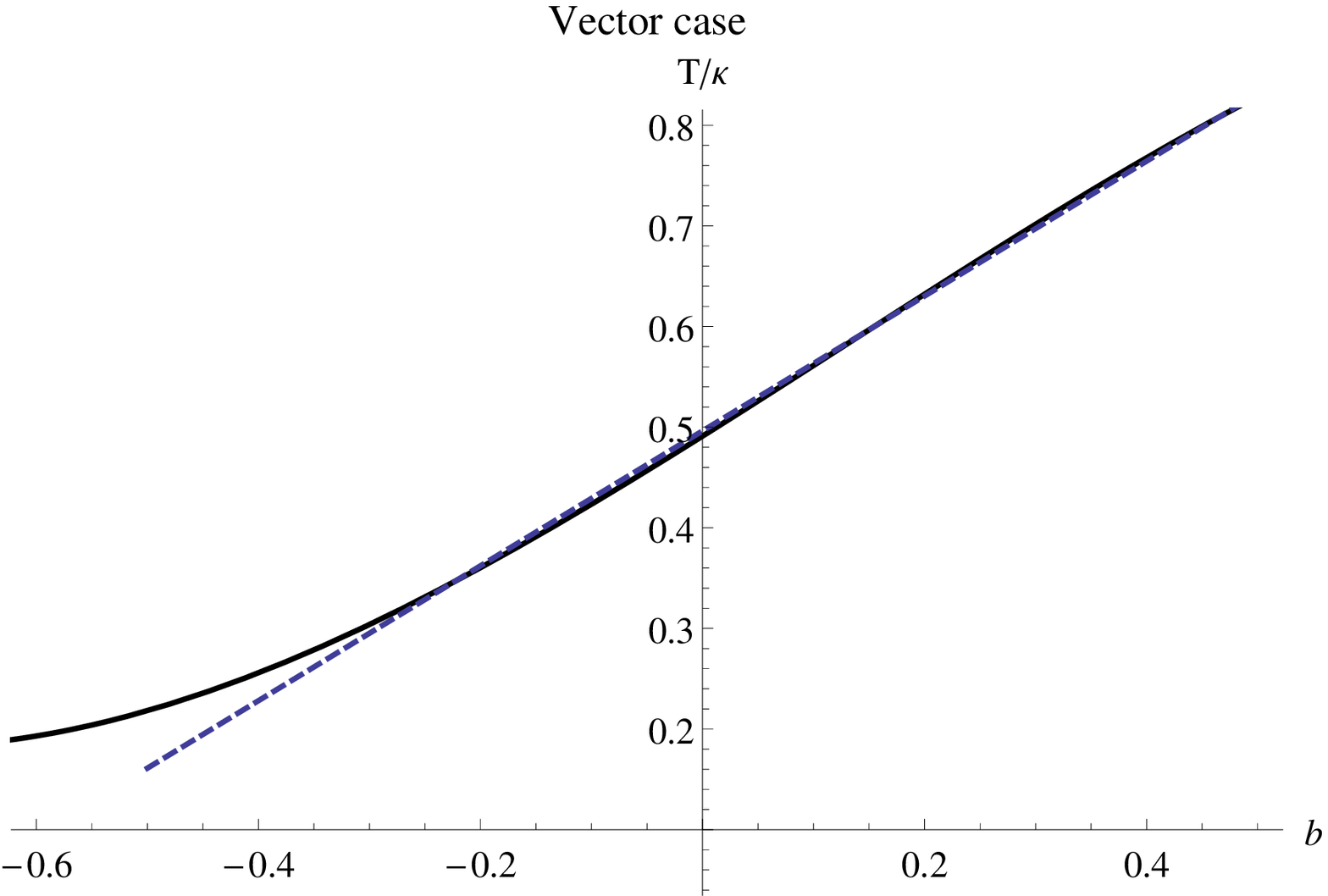} \includegraphics[scale=0.45]{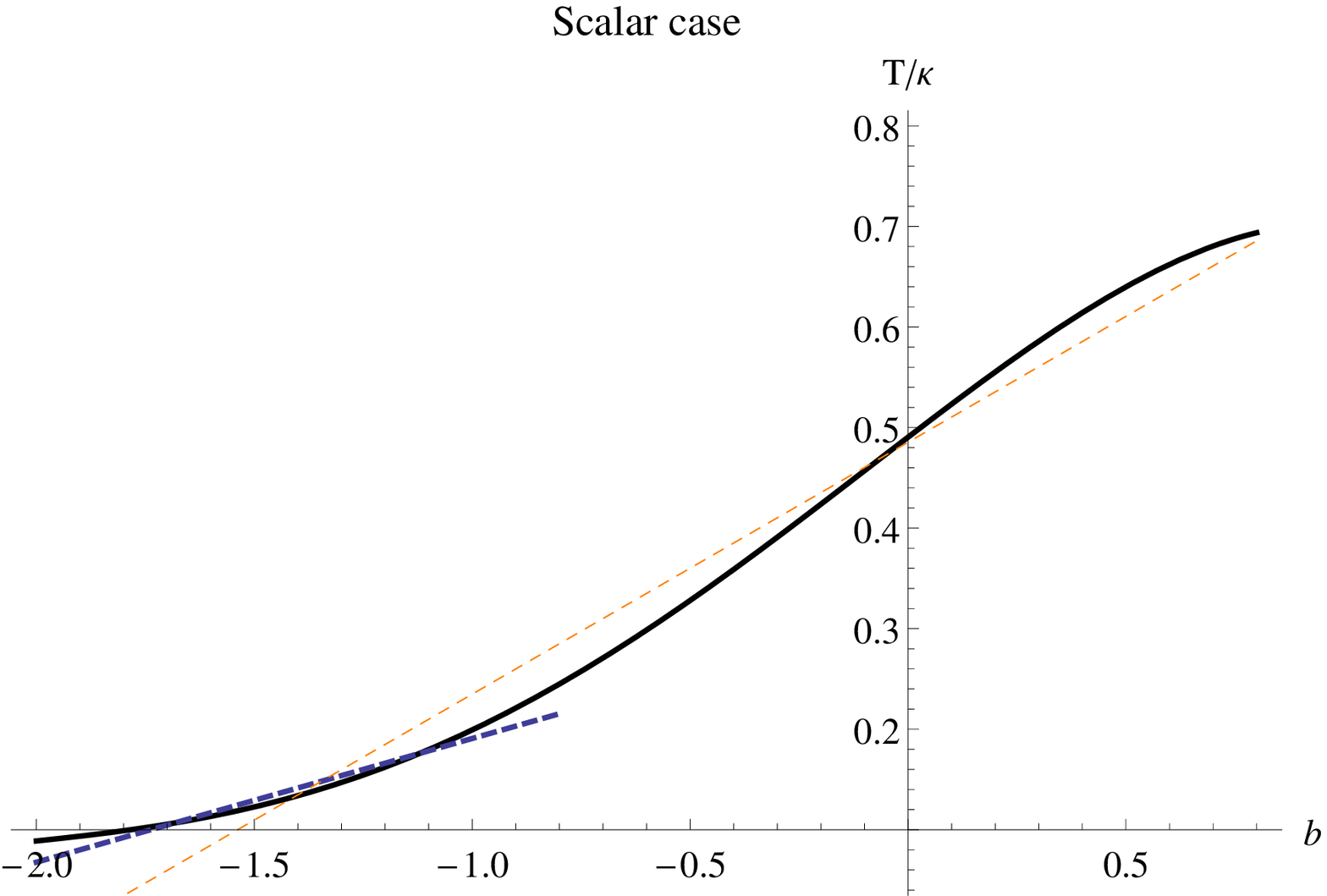}	
	\caption{\label{GSWtemp} The deconfinement temperature in GSW model as a function of the intercept parameter $b$. The blue 
	dashed lines are the linear interpolation functions given in Eqn.~(\ref{Tofb}).}
\end{figure}

In the GSW case the following numerical approximations are valid for the values of $b$ corresponding to
phenomenological spectra (to be discussed in Section~\ref{experiment}), see Fig.~(\ref{GSWtemp})
\be\label{Tofb}
T^{J=1}_{GSW}/\kappa \simeq 0.670\cdot b + 0.496,\quad 
T^{J=0}_{GSW}/\kappa \simeq 0.123\cdot b +0.314.
\ee
Again, we postpone the discussion of numerics to Section \ref{holo_T}.

An interesting but speculative feature of the SW model is the usage of the inverse dilaton profile \cite{inverse_dilaton},
that means the case $f^2(z)=e^{\kappa^2z^2}$. Among other things this choice modifies the intercept of the particle spectra.
In the context of AdS/QCD study of glueballs the inverse dilaton option is sometimes considered. For instance, in \cite{BoschiFilho:2012xr}
a modified SW model is investigated and the negative dilaton supposedly provides the best fit for the scalar glueball masses.
From the point of view of thermodynamic properties, we have to argue that the inverse dilaton puts the theory in the permanently deconfined
(black hole) phase. It is straightforward to reproduce the previous calculations for this case and see that
$\Delta V$ is always negative and no phase transition is possible.

\section{Isospectrality}
\label{isospectrality}
All AdS/QCD outputs being at least indirectly connected to the spectra of the resonances, it is worth mentioning a recent proposal 
to modify the holographic models preserving the spectrum form. In \cite{Vega_Cabrera}, one finds such a study
featuring a family of dilatons related to the usual SW one. The family is achieved through the notion of the 
isospectral potentials.

We propose to investigate the way the predictions of the deconfinement temperature  vary for different members of a dilaton
family. In particular, we will adopt the method of \cite{Vega_Cabrera} to the massless $5D$ fields 
subject to the EOM~(\ref{jEOM}) and derive an extension to the case of GSW.

It is straightforward under a certain change of variables $\Phi(q,z)=f(z)^{-1} z^{(3-2J)/2} \Psi(q,z)$
to rewrite  Eqn.~(\ref{jEOM}) for the KK $z$-profiles $\psi_n(z)$ in a Schr\"{o}dinger form
\be\label{Schrod}
-\psi''_n(z)+\widehat{\mathcal V}(z) \psi_n(z) =M^2(n)\psi_n(z).
\ee
Here $\widehat{\mathcal V} (z)$ is the Schr\"{o}dinger potential for a model with unspecified dilaton function $f^2(z)$. We further add a subscript $J$ because the potential is generally dependent on the spin parameter $J$ as follows,
\be
\widehat{\mathcal V}_{J} (z)=\frac{(3-2J)(5-2J)}{4z^2}+\frac{f''(z)}{f(z)}-\frac{3-2J}z\frac{f'(z)}{f(z)}.
\ee
For the dilaton profile of the GSW model, it is given by
\be
\mathcal V_{J}(z)=\frac{(3-2J)(5-2J)}{4z^2}+\kappa^4z^2+2\kappa^2(1-J+2b).
\ee
A particular form of the Schr\"{o}dinger potential defines the eigenvalues of Eqn.~(\ref{Schrod}) and hence
the mass spectrum $M(n)$. In the case of (G)SW models it is a potential similar to the one that appears
when considering the radial part of the wave function of a $2D$ harmonic oscillator system. The eigenvalues
being known, we simultaneously gain the spectrum of Eqn.~(\ref{GSW_sp}).

According to \cite{Vega_Cabrera} and the references therein,
the following isospectral transformation between $\mathcal V_J(z)$ and  $\widehat{\mathcal V}_J(z)$ exists
\be\label{Darboux}
\widehat{\mathcal V}_J(z) = \mathcal V_J(z) -2\frac{d^2}{dz^2}\ln[I_J(z)+\lambda].
\ee
This technique allows us to generate a family of dilaton functions $f(z)$ appearing in $\widehat{\mathcal V}_J(z)$,
each member assigned to the value of the parameter $\lambda$ (we assume $0<\lambda<\infty$). The case of $\lambda=\infty$ corresponds to the original $\mathcal V_J(z)$.
The function $I_J(z)$ is defined through the ground eigenstate of $\mathcal V_J$, $\psi_0$, and is given by
\be
I_J(z)\equiv\int\limits_0^z\psi_0^2(z')dz'=1-\frac{\Gamma(|2-J|+1,\kappa^2z^2)}{\Gamma(|2-J|+1)}.
\ee
Different $\lambda$ provide slightly different forms of the potential, but the eigenvalues of Eqn.~(\ref{Schrod})
and, hence, the spectrum remain the same.
The given trick is well known in Supersymmetric Quantum Mechanics \cite{SUSY_QM}
and represents a feature of one-dimensional potentials: The discrete spectrum of normalizable modes does not fix the potential, there is an infinite family of such isospectral potentials which are related by the transformations
of the kind~(\ref{Darboux}). In this construction the parameter $\lambda$ has no direct physical meaning and just reflects this specific spectral 'symmetry' in the 'space' of one-dimensional potentials. 

We restrict the model to deviations only in the exponential factor, {\it i.e.} the isospectral
profiles of a type $f^2(z)=\exp(-\chi(z))U^2(b,J-1;\kappa^2z^2)$, with the asymptotes fixed $\chi(z\rightarrow0)=\chi(z\rightarrow\infty)=\kappa^2z^2$.
Introducing an argument $t=\kappa^2z^2$, we may define the family of $\lambda$-dependent equations as follows:
\be
\frac{tf''(t)+(J-1)f'(t)}{f(t)}-b
=\frac{t+2(1-J)}4-\left(\frac{d}{dt}+2t\frac{d^2}{dt^2}\right)\ln \left(I_J(t)+\lambda\right).
\ee

For a given $b$, $J$ and $\lambda$ an interpolating function for $\chi(\lambda, z)$ may be numerically found and substituted afterwards
 inside $f^2(z)$ in the $\Delta V$ equation:
\begin{equation}
\Delta V_{GSW} = \frac{\pi R^3}{k_g z_h}\left[\frac{U^2(b,J-1,0)}{2}-f^2(\kappa^2z_h^2)-
\kappa^2z_h^2\left.\frac{d f^2(t)}{dt}\right|_{t=\kappa^2z_h^2}-(\kappa z_h)^4\int\limits_{\kappa^2z_h^2}^{\infty}dtt^{-1} \frac{d^2 f^2(t)}{dt^2}\right].
\end{equation}	
As a family parameter $\lambda$ defines different $\chi(\lambda, z)$, the solutions of this equation and hence the critical temperature may vary.
Are these deviations of a large scale, or does isospectrality preserve isothermality in general?
If not, could  we choose a correct family member with a 'physical' value of the formal parameter $\lambda$?  
May be there are some specific fits to (G)SW that provide more stable results than others.
We suggest to explore these options in Section \ref{holo_T}.

\section{Lattice and experiment}\label{experiment}
\subsection{Deconfinement temperature}
The deconfinement transition in the QCD matter is a complicated and largely not-understood process.
Basically, we presume that with the temperature growth the description in terms of hadronic states becomes worse
and worse until finally one should turn to consider the hadron matter
as the quark-gluon plasma. One would like to define some critical parameter $T_c$ at which the change happens.
Further, one may turn to the experimental data on heavy ion collisions, introduce the (model dependent)
way to extract the particular information and achieve the constant value of $T_c \simeq 160$~MeV ({\it e.g.} see \cite{Andronic:2009}).

Alternatively, a lot of investigations on lattice have been performed to study the problem. They leave no doubt to
the fact that the deconfinement happens smoothly with temperature and represents rather a crossover than a 
phase transition \cite{Aoki:2006we}, \cite{Borsanyi:2010bp}.
Hence, the notion of the critical parameter is substituted with the pseudo-critical temperature, the precise
determination of which is subject to the method and may vary in almost $20\%$ range.
The results of lattice simulations with physical quarks \cite{Borsanyi:2010bp} provide the values
$T_c \sim 150-170$~MeV.

On the other hand, pure Yang--Mills theories (with no dynamical quarks at all) exhibit a somewhat similar phase structure. The confined 
phase corresponds to the bound glueball states, while the deconfined -- to the gluon plasma.
Here the deconfinement happens as a true phase transition, and it is of a first order for $SU(3)$.
Turning to the lattice studies of $SU(3)$ theory \cite{Boyd:1996bx},\cite{Iwasaki:1996ca} 
we encounter $T_c \sim 260-270$~MeV.
Lattice QCD simulations with non-dynamical quarks in the limit $N_c\rightarrow\infty$ describe a similar physical
system and the prediction there is $T_c/\sqrt{\sigma}=0.5949 + 0.458/N_c^2$  \cite{Lucini_temper2012}.
The temperature, as well as other lattice outputs, is generically measured in terms of the dimensional
quantity -- the string tension $\sigma$. With the standard choice $\sqrt{\sigma}=420$~MeV we get
$T_c \sim250$~MeV.

The temperature estimations of the last paragraph should be the most relatable to the holographic predictions as
it is defined in the leading $N_c$ order and as we associate the deconfinement holographic transition with the first order 
Hawking -- Page phase transition. Nevertheless, it is an often practice to compare AdS/QCD predictions
with the `real QCD' results of $150-170$~MeV. The 'pro' arguments here are that the strict $N_c=\infty$ limit is almost
always softened to achieve real phenomenology, there is a relatively good description of hadron resonances in AdS/QCD, and
the critical parameter turns to be a satisfactory approximation of the pseudo-critical lattice one (based on particular numerical fits, 
some controversy of which is discussed in Section \ref{holo_T}).

\subsection{Assumed $0^{++}$ glueball states}
\subsubsection{Lattice calculations}
\begin{figure}[t]
	\includegraphics[scale=0.5]{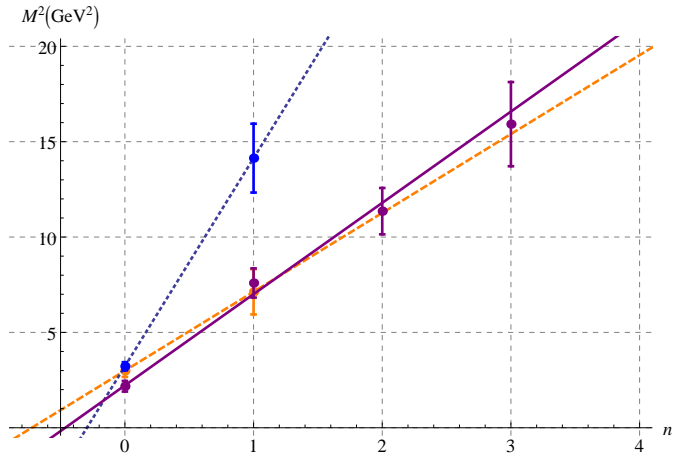}
	\caption{\label{glue_traj} The radial Regge trajectories of glueball states:
		dashed -- M $\&$ P \cite{Morningstar:1999},  solid -- Meyer \cite{Meyer_PHD}, 
		dotted -- unquenched \cite{Gregory_unquenched_2012}.}
\end{figure}
The lattice studies of glueball states and investigation of their masses were mostly performed in the quenched
approximation, {\it i.e.} in the pure gluon theory. Further, we quote several works simulating $SU(3)$ on lattice:
\begin{itemize}
 \item Morningstar and Peardon \cite{Morningstar:1999} in 1999 report on $0^{++}$ (with a mass of $1730(50)(80)$~MeV)
and  $0^{++*}$ states which approximately follow the radial Regge trajectory $m^2=4\cdot (1017\ \text{MeV})^2 \cdot (n + 0.72)$.
\item  Meyer in \cite{Meyer_PHD} finds a ground state at $1475(30)(65)$~MeV and three radial excitations which belong nicely 
 to a linear trajectory $m^2=4\cdot (1094\ \text{MeV})^2 \cdot (n + 0.46)$; see Fig.~\ref{glue_traj}.
\item Chen {\it et al.} in \cite{Chen:2005} discuss the $0^{++}$ state
that has the mass of $1710(50)(80)$~MeV.
\end{itemize}

It is noted in \cite{Gregory_unquenched_2012} that the possible source of
a systematic difference in the results of Meyer 
 with respect to the ones of  Morningstar and Peardon and Chen {\it et al.} for the ground state is the usage of the string
tension $\sigma$ versus the hadronic scale parameter $r_0$ to determine the lattice spacing.
However, one can notice from  Fig.~\ref{glue_traj} that the full trajectory of \cite{Meyer_PHD} is in a general accordance with an interpolation of
\cite{Morningstar:1999}.

The large $N_c$ limit is sometimes considered as well, as in practice the results for not so high group degree, say 
$SU(8)$, do not alter much from the infinite $N_c$ extrapolation.
We find the results of \cite{Lucini_inf_2004} the most trustworthy and quote them in terms of $\sigma$:
$\frac{m^2}{\sigma}=4\cdot 2.55 \cdot (n + 0.42)$ (they are a revised version of a widely quoted \cite{Lucini_inf_2001}).
In \cite{Meyer_PHD} the results for $SU(8)$ are claimed to be a valid approximation of
$N_c \rightarrow \infty$ limit and are given by: $\frac{m^2}{\sigma}=4\cdot 1.67 \cdot (n + 0.99)$.
 
The unquenched approximation provides another interesting viewpoint. Though earlier it was supposed that
glueball masses remain almost the same or get a $20 - 40 \%$ suppression with respect to the quenched results,
the authors of the most recent work on this subject
\cite{Gregory_unquenched_2012} report on the ground state of $1795(60)$~MeV, that is greater than any aforementioned value.
Together with the first excitation the unquenched method of \cite{Gregory_unquenched_2012}
provides a trajectory $m^2=4\cdot (1652\ \text{MeV})^2 \cdot (n + 0.30)$. Interestingly, the slope is much
steeper than in the quenched approximation due to the $0^{++*}$ state of \cite{Gregory_unquenched_2012}
being in the range of $0^{++***}$ of \cite{Meyer_PHD}, as is clear from Fig.~\ref{glue_traj}.

\subsubsection{Candidate among $f_0(1370)$, $f_0(1500)$, $f_0(1710)$}
Having established from the lattice a region of masses where the scalar glueball may be encountered,
$1.5 - 1.7$~GeV, one looks there for the non-strange mesons with the same quantum numbers that do
not fit into $q\bar q$ nonets.
The most often regarded hypothesis emerges then: $f_0(1370)$, $f_0(1500)$, $f_0(1710)$
states are a mixture of $\bar u u + \bar d d$, $\bar s s$ and glueball modes.

However, a common viewpoint on the degree of the mixing, or on what state could
be mostly glueballic, does not exist. 
For a variety of possibilities, in broader mass ranges as well, see, for instance, \cite{Ochs:2013}.
An attempt to separate in the set of known $f_0$'s the radial Regge trajectory of mesonic states from the glueballs from the point of
view of rotated closed strings in the holographic background was made by the authors of \cite{Sonnenschein:2015zaa}.
Though various interesting fits to different glueball candidates are presented there, there is no final conclusion
but that ``an extension of experimental data on the spectrum of flavorless hadrons is needed''  \cite{Sonnenschein:2015zaa}.
In this paper, we do not try to cover all the options and will focus on the two main ones.

The first model, rather strongly advocated in Particle Data \cite{PDG}, is assuming 
that $f_0(1500)$ has the largest glueball component. $f_0(1370)$ consists mostly of the up and down quarks
and $f_0(1710)$ of the strange ones.
For a typical example, we further take the results of \cite{Close:2005vf}, in which two fits
to existing data are provided: Fit I gives  $m_{gl} = 1464\pm47\ \text{MeV}$, and
Fit II -- $m_{gl} = 1519\pm41\ \text{MeV}$.

The second model proposes exchanging the roles of $f_0(1500)$ and $f_0(1710)$ and, suggesting that the latter 
has a predominant glueball component, gives $m_{gl} \sim 1665\ \text{MeV}$ \cite{Cheng:2006hu}.
 A revised version of this model \cite{Cheng:2015} provides $m_{gl} = 1674 \pm 14\ \text{MeV}$.
 
Shortly, the first model is mostly supported by the fact that $f_0(1500)$ are not encountered in $\gamma\gamma$
reactions. In the same time the authors of  \cite{Cheng:2015} argue that this fact does not necessarily imply a
large glueball component there and advocate the $f_0(1710)$ option from the point of $J/\psi$ decays.
$f_0(1710)$ is also concluded to be a ground glueball state in some theoretical works
(\cite{Janowski:2014ppa}, \cite{Brunner:2015}).

\subsection{Vector mesons}
\begin{figure}[t]
	\includegraphics[scale=0.7]{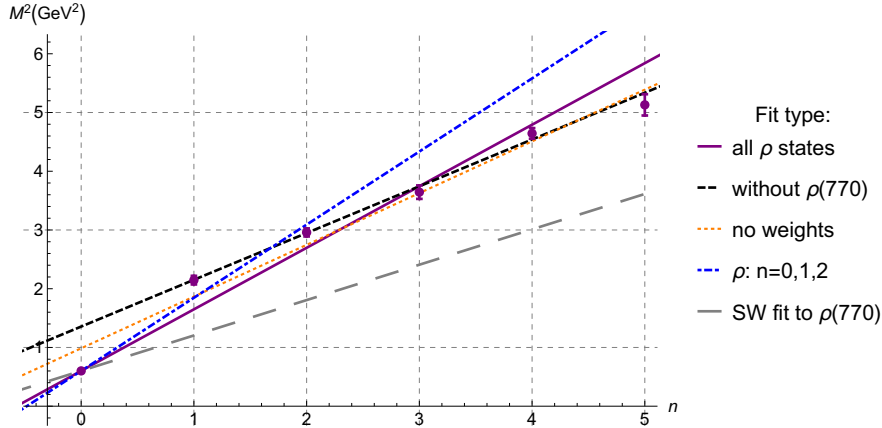}
	\caption{\label{AllRho} The possibilities for the radial Regge trajectories of the $\rho$ meson.
	}
\end{figure}
The $\rho$ and $\omega$ mesons, despite being seemingly well defined particles,
pose certain trouble for one aspiring to explore the spectrum of their radial excitations in 
much detail: the ground state lying notably below the linear trajectory, broad resonances, {\it etc.}
See \cite{estim_deconf} for some discussion on these issues.

The ground state poses the main problem in the present study as, on one hand, it has the most established
mass value, being most useful for the numerical estimations, but on the other hand it is
deviating significantly from the linear trajectory of the higher radial excitations, see Fig.~\ref{AllRho}. The latter
fact questions and obscures the usage of these trajectories in the SW-like models.
It should be mentioned that there are many specific modifications of the SW model which lead to certain non-linearities of radial trajectories -- dynamical \cite{Bartz:2014oba}
and non-dynamical \cite{Afonin_finite},\cite{Cui2016} modifications of the dilaton and AdS metric as well as SW models containing a cut-off of the holographic coordinate \cite{Afonin:2011ff}.
Nevertheless, we try to make several realistic hypotheses with linear trajectories in the following section.

\section{Holographic predictions for $T_c$}\label{holo_T}
A general disclaimer for this section is that we consider all numerical results at the level of an estimation.
The appearing error bars are only due to the uncertainty of the experimental or lattice determination of
the input parameters.
The holographic apparatus can only claim numerical validity on the semi-quantitative level and we cannot 
wholly estimate the theoretical error.
Though for the well-known masses of the vector mesons the experimental error is definitely small and not comparable 
with the theoretical one, the lattice results are not that precise, and we find it useful to provide
the minimal possible error that could be present.
\subsection{Deconfinement temperature from the vector channel}
\begin{table}
\begin{tabular}{|*{4}{c|}}  
	\hline
	Fit $\&$ Method&  \multicolumn{3}{c|}{$T_c$ (MeV)}\\
	\cline{2-4}
	 &$\rho$ meson & $\omega$ meson & 'universal' trajectory\\
	 \hline
	lightest state $\&$ HW & $122.03\pm0.04$  & $123.19\pm0.02$&'$168.1\pm0.7$' \\
	lightest state $\&$ SW & $190.61\pm0.06$  & $192.43\pm0.03$ & $262.5\pm1.2$\\
	$n=0,1,2\ \&$ GSW & $118.3\pm2.7$  & $121.4\pm4.0$ & --\\
	\hline
\end{tabular}
\caption{Standard vector meson predictions for the deconfinement temperature in HW and (G)SW.}
\label{standard_T}
\end{table}

Eventually, we are going to discuss the numerical holographic predictions on $T_c$ and 
start with the vector case, $J=1$. 

In Table~\ref{standard_T} we cover the results some of which were previously given
in  \cite{Herzog:2006} and \cite{estim_deconf}. The first two lines represent the classical AdS/QCD fit
with the lightest state (usually a $\rho$-meson), 
while the third -- the fit to the tower of the established resonances (the ground state and the first and second radial excitations).
It is evident from Fig.~\ref{AllRho} that the SW fit to the ground state does not follow the real trajectory
of the $\rho$ excitations. Neither does the HW spectrum, that goes as $M(0)\{1; 2.3; 3.6; ...\}$ featuring much
faster growth than the actual $\rho$ spectrum $M(\rho)\{1; 1.9; 2.0; ...\}$. 
One would expect the inclusion of higher excitations to help with the spectrum description,
but it does not happen due to the high weight of the ground state in the fit.
For instance, the $n=0,1,2$ trajectory deviates significantly from the further states and 
provides the temperature estimation of $118$~MeV that does not seem satisfying. The same can be attributed to
the $n=0,1,2,3,4$ fit resulting in $T_c \simeq 125$~MeV.
Thus, all the $T_c$ predictions of Table~\ref{standard_T} from $\rho$ and $\omega$ fits rely on the 
non-standard position of the lightest state and do not come close to any lattice estimate.

One way around this is to include equal weights in the fit as was performed in \cite{estim_deconf} in hope of getting 
an average fit that can bring better results for the temperature
estimations. Indeed, this strategy resulted in $T_c \simeq 150$~MeV for $n=0,1,2$ trajectories.

Another option is presented in the third column of Table~\ref{standard_T}. As argued in \cite{estim_deconf} we find the fit with 'universal' slope to be the most successful.
There are some phenomenological reasons for the universality of the slope for various light non-strange
mesons as noticed in~\cite{ani,bugg} and further advocated in many studies~\cite{KLEMPT20071, *ShifVainsht, *Afonin:2006vi,
*Afonin:2007jd,*Afonin:2007mj, *Afonin:2007sv, *Afonin:2007bm, *Li:789523, *Masjuan2012, *AfoninPusenkovPRD, *AfoninPusenkovletter, *Masjuan:2014sua}.
As well the universality of this kind is in concordance with
the hypothesis (inspired by the hadron string models) that
the slope is mainly determined by the gluodynamics.
From the holographic (G)SW model point of view the universality of the slope means the universality of
the dilaton profile for the particles of any spin. Moreover, the use of
the universal slope ($\kappa=534\ \text{MeV}$ \cite{bugg}) provides 
a unique $T_c\simeq 260$~MeV. This result being in range of lattice predictions with  non-dynamical quarks allows us
to claim that the fit defined from a gluodynamic insight provides the deconfinement temperature expected from gluodynamics.
It was highlighted in the Introduction and could be seen below that this fit corresponds to selecting $f_0(1500)$ meson 
as the lightest scalar glueball candidate.

\begin{table}[t]
	\begin{center}
		\begin{tabular}{|*{4}{c|}} 
			\hline
			$\lambda$ & \multicolumn{2}{c|}{$T_{SW}$ (MeV)}& $T_{GSW}$ (MeV) \\
			\cline{2-4}
			&'universal' trajectory & lightest $\rho$ meson& $\rho:\ n=1,2,3,4,5$ \cite{PDG}\\
			& $m^2= 4\cdot (534\ \text{MeV})^2\cdot(n+1)$& $m^2= 4\cdot (388\ \text{MeV})^2\cdot(n+1)$&$m^2= 4\cdot (446\ \text{MeV})^2\cdot(n+1.7) $ \\
			\hline
			$100$ & $261.1\pm1.2$  &$189.66\pm0.06$ &$422.5\pm34.2$ \\
			$20$ &   $256.1\pm1.1$  &$185.97\pm0.06$ &$419.0\pm34.5$ \\
			$1$ &  $194.9\pm0.9$   &$141.54\pm0.05$ &$339.2\pm39.2$\\
			$0.1$ &  $159.1\pm0.7$  & $115.53\pm0.04$&$189.2\pm11.4$ \\
			$0.01$ & $154.3\pm0.7$  & $112.03\pm0.04$&$172.4\pm8.0$ \\
			\hline
		\end{tabular}
	\end{center}
	\caption{Dependence of the deconfinement temperature estimations from vector meson fits in the (G)SW model on the isospectral
		parameter $\lambda$.}
	\label{v_isosp}
\end{table}

Now, let us consider the families of isospectral dilatons and their impact on the value of $T_c$,
that is provided in Table~\ref{v_isosp}.  It is clear that for vector mesons isospectrality does not
come together with isothermality. There are two ways to treat this problem. First is to say that, taking
into account the scale of changes in $T_c$ with different $\lambda$, none of the results on the deconfinement
temperature in this model are trustworthy. Another is to consider that if holography exists in nature,
there could be various reasons for $\lambda$ to take a particular 'true' value, not essentially the original
$\lambda=\infty$. What other physical processes affect the particular choice of $\lambda$ corresponding
to the 'true' $T_c$ may be a subject of future investigations.

We have chosen several noteworthy examples in Table~\ref{v_isosp}. For the 'universal' trajectory we see a transition
of the $T_c$ from the region predicted on lattice with non-dynamical quarks (about $260$~MeV)
to the one of the unquenched approximation ($150-170$~MeV).
For the standard SW model we can get a viable prediction at some finite $\lambda$. For the GSW we take a specific
fit trying to specify the role of $\rho(770)$  -- we just exclude the ground state (see Fig.~\ref{AllRho}). Though at $\lambda=\infty$ this trajectory
provides a very  large and unnatural $T_c$,  we can go down to the expected values 
at finite $\lambda$. At least this consideration allows us to avoid the
complete exclusion of the higher $\rho$ trajectory hypothesis
following the criterion of viable $T_c$ predictions.

\subsection{Deconfinement temperature from the glueball channel}
\begin{table}
\begin{tabular}{|*{6}{c|}}  
	\toprule\hline
	Fit & $m_{gl}$ (MeV) & $T_{HW}$ (MeV) & \multicolumn{3}{c|}{$T_{SW}$ (MeV)}\\
	\cmidrule{4-6}\cline{4-6}
	 & & & $\lambda =\infty$ & $\lambda = 1 $ & $\lambda =0.1$ \\
	\midrule\hline
	M $\&$ P \cite{Morningstar:1999} & 1730(100) & 171(10) & 301(17) & 253(15) & 173(10) \\
	Meyer \cite{Meyer_PHD} &1475(75) & 146(7)& 256(13)& 215(11) & 147(8)\\
	Chen {\it et al.} \cite{Chen:2005} &1710(95) & 169(9)& 297(17)& 250(14) & 171(10)\\
	\hline
	Large $N_c$ \cite{Lucini_inf_2004} &1455(70) & 144(7)& 253(12)& 212(10) & 145(7)\\
	\hline
	Unquenched \cite{Gregory_unquenched_2012} &1795(60) & 177(6)& 312(10)& 262(9) & 179(6)\\
	\hline
	\multirow{2}{*}{$f_0(1500)$ meson \cite{Close:2005vf}} & 1464(47) & 145(5)& 255(8) & 214(7) & 146(5) \\
					  &1519(41)& 150(4) & 264(7)& 222(6) & 152(4)\\
	$f_0(1710)$ meson \cite{Cheng:2015} & 1674(14)& 165(1)& 291(2) & 244(2) & 167(1)\\
	\bottomrule\hline
\end{tabular}
\caption{\label{1gl-tab} One glueball state predictions in HW and SW models.}
\end{table}

To begin with scalar glueball estimations of $T_c$, we propose to combine lattice results 
in $N_c \rightarrow \infty$ limit for the relation of  $T_c$ to $m_{gl}$, where $m_{gl}$ is the mass of the $0^{++}$ state.
Both quantities are measured on lattice in terms of $\sqrt\sigma$, so such fraction provides a $\sqrt\sigma$ independent result.
From \cite{Lucini_temper2012} and \cite{Lucini_inf_2004} we get
\be
\left.\frac{T_c}{m_{gl}}\right|_{lat}=0.1799.
\ee
While the HW and SW predictions are
\be\label{T_wosigma}
\left.\frac{T_c}{m_{gl}}\right|_{HW}=0.0988, \quad \left.\frac{T_c}{m_{gl}}\right|_{SW}=0.1739.
\ee
The SW result is rather close, much better than the one achieved in the models of Improved holographic QCD:
$\left.\frac{T_c}{m_{gl}}\right|_{IhQCD}=0.167$ \cite{Gursoy:2010fj}.
If both states reported in \cite{Lucini_inf_2004} are considered and supposed to lie on GSW trajectory, 
we get $\frac{T_c}{\sqrt{\sigma}}=0.2988$ to be compared with $0.5949$ of \cite{Lucini_temper2012}. 
This does not seem satisfying, but remarkably isospectral methods change this result only 
in the last digit.

Utilizing the formulas of Eqn.~(\ref{T_wosigma}) and taking the $0^{++}$ glueball mass
in the physical scale we produce the variety of results
of Table~\ref{1gl-tab}. The isospectrality is considered for SW model and features similar behaviour to what
we have seen in the vector case. The HW results converge to a region of $150-170$ MeV, close to the lattice
with dynamical quarks. The SW predictions are a bit larger than the quenched lattice ones, though it is possible
to get lower within the isospectral family. Both fits of the hypothesis of $f_0(1500)$ being a mostly 
glueball state predict $\sim 260$~MeV in SW and generally are close to the fit of the 'universal' vector trajectory.
The large $N_c$ fit provides a similar number $\sim 250$~MeV in SW, showing nice concordance with large $N_c$ predictions from the lattice.

\begin{table}
\begin{tabular}{|*{7}{c|}}  
	\toprule\hline
	Fit &$\sqrt{\sigma}$ or $r_0^{-1}$  &\multicolumn{2}{c|}{$m^2=4\kappa^2 (n + 2+b)$} &  \multicolumn{3}{c|}{$T_{GSW}$ (MeV)}\\
	\cline{3-7}
	&(MeV) &$\kappa$ (MeV) &b & $\lambda =\infty$ & $\lambda = 1 $ & $\lambda =0.1$ \\
	\midrule\hline
	M $\&$ P \cite{Morningstar:1999}& $410$ & $1017(151)$&$-1.28(0.23)$ & $153.6(39.2)$ & $151.4(36.1)$ & $149.9(34.1)$ \\
	Meyer \cite{Meyer_PHD} & 440 & $1094(49)$ &$-1.54(0.07)$ & $132.5(9.7)$ & $132.3(9.5)$ & $132.1(9.4)$ \\
	Unquenched \cite{Gregory_unquenched_2012}& 420& $1652(138)$ & $-1.71(0.05)$& $174.6(16.4)$& $174.6(16.4)$ & $174.5(16.4)$\\
	& & & & & &\\
	Large $N_c$  \cite{Lucini_inf_2004}& 440 & 1120(88)& $-1.58(0.08)$ & 131.3(13.5) & 131.1(13.4)& 131.0(13.3) \\
	Large $N_c$ \cite{Meyer_PHD}& 440 & 735(121)&$-1.00(0.35)$ & 142.8(55.3) & 134.8(43.7)& 129.7(36.6) \\
	\bottomrule\hline
\end{tabular}
\caption{\label{tower-gl-tab} Predictions of $T_c$ from different fits for glueball towers in GSW model with isospectrality.}
\end{table}

Next, in Table~\ref{tower-gl-tab}, we take the full spectra of radial excitations and 
define $T_c$ through GSW formulas to achieve astonishing isothermality.
The error bars here are much larger as the higher glueball excitations are not that well-measured
and usually only the masses of $0^{++}$ and $0^{++*}$ states are  calculated.
First, that does not allow us to be sure of the validity of linear Regge assumption for the radial excitations
of the scalar glueballs. Second, the slope error may get rather significant. However, the only work reporting
on more than two states, \cite{Meyer_PHD}, gives a more well-defined trajectory
(and rather linear, see Fig.~\ref{glue_traj}). The $T_c$ prediction
there is rather low, but that is due to the fact that the ground state is calculated to have a mass $\sim 1.5$~GeV,
which may be considered systematically lower than other lattice predictions for the masses.
Altogether, we find all $T_c$ of Table~\ref{tower-gl-tab} to be rather close to the results of unquenched
lattice predictions for the deconfinement temperature,
especially it is significant to have this agreement for the unquenched fit of \cite{Gregory_unquenched_2012}.

\section{Concluding discussions}\label{conclusions}
A general motivation of this paper is to raise awareness that, while making AdS/QCD calculations,
one should be concerned from where and for what reasons one takes the input parameters.

We can easily assume an occurrence of different dilaton pre-factors in the gravity and matter parts of the $5D$ action.
Moreover, for any scalar, vector and tensor operators in a given strongly interacting theory
we may construct a $5D$ Lagrangian with a suitable $5D$ field following the AdS/CFT dictionary.
If we want to consider a holographic description involving all this physics (at the level the
bottom-up holography can manage), we should construct a general Lagrangian of the type
\be
S_{5D} = \int d^5x\sqrt{-g} \left(f^2_{gr}(z)\mathcal L_{gravity} 
+ \sum_j f^2_{sc\ j}(z)\mathcal L_{scalar\ j} + \sum_j f^2_{v\ j}(z)\mathcal L_{vector\ j} + ... \right).
\ee
Here, $f^2(z)$  schematically means HW, (G)SW or any other $z$-dependent pre-factor induced by extra-dimensional dynamics
together with the restrictions on the $z$ interval. We could generally assume that all $f^2_j(z)$ are different.
This option is not theoretically well-motivated as the dilaton field is a certain entity of the string theory,
and, for instance, in the models of dynamically generated SW it  comes as a unique solution of Einstein equations
emerging from a particular graviton--dilaton $5D$ action (see the works \cite{Li:2013oda} and \cite{Bartz:2018nzn} for the interesting results concerning glueballs in the models of this type). 
But phenomenologically, this only seems natural, and, for example, the case of
$\kappa_{sc}=\kappa_v$ in a standard SW is a feature that may be or may be not relevant in the physical model we want to study.

Then why should the gravity dilaton be coincident with some other? Generally, from the bottom-up point of view only, it should not and may be considered as another free parameter.
Nevertheless, it is our main objective to determine the region to which the value of $T_c$ belongs,
and the model predictability cannot be given up so easily.

We find no other option than to try to consider several possibilities and determine the 'correct' one (if it exists)
not only for the conceptual reasons but also for the phenomenological ones. An expense of free parameters not being
a feature we require, we utilize the least fine-tuned holographic frameworks: HW, SW and GSW. 
For the matter content, we make the traditional choice of vector mesons and the one related to the similarity of the $N_c$ scaling
of $4D$ gluodynamics and $5D$ gravitational action, {\it i.e.} $0^{++}$ glueballs and its radial excitations.
For an optional check in SW-like models, we propose aspiring to constant temperature predictions in an isospectral family.

For a minimal option, we argue once more on the relevance of the idea of the 'universal' slope value for the 
radial trajectories of light non-strange mesons. That means a fixation of
the dilaton parameter in $f^2_{gr}(z)=e^{-\kappa^2z^2}$ to $\kappa\simeq530$~MeV, that coincides with the case of $f_0(1500)$ being predominantly a
$0^{++}$ glueball. 
This results in $T_c\simeq 260$ MeV corresponding exceedingly well to the lattice results for pure $SU(3)$
and large $N_c$ limit. Both heuristics and numbers are in favour of this hypothesis.
Moreover, this is the least model dependent result in our considerations.
The lack of isothermality in the isospectral family of this dilaton seems to be the only downside. 
	
The simplest models, HW and SW, feature a rough fitting to the radial trajectories
as a whole (especially in HW), and it is convenient to restrict them to reproduce the masses of the ground states.
For scalar glueballs, it is even more essential, as the ground states are much better identified on lattice than their excitations.
We find that for $0^{++}$ glueball candidates (from lattice or from the identification with some $f_0$),
which masses are limited within the range of $1.5 - 1.7$ GeV, the deconfinement temperature lies in a range of
$260-290$ MeV in SW and $150-170$ MeV in HW. For the $\rho$ or $\omega$ mesons we get $\sim 190$~MeV in SW 
and $\sim 120$~MeV in HW. Clearly, the results of HW and SW differ, though for the glueballs they appear to coincide 
with lattice expectations in different regimes. That is difficult to interpret, though we can notice that going through the 
isospectral family we can connect these separated regions (perhaps, in a way, mimicking
quark masses becoming physical on lattice).

The generalized SW models provide a more accurate description of excited spectrum. We studied the predictions from the various spectra of vector mesons
in \cite{estim_deconf}. There, we have seen that different spectra may result in various predictions for $T_c$, not providing
a clear way to select the best fit. Here, we find out that additionally these temperature estimates vary a lot as we go through
the isospectral family. In this work, we performed a similar analysis for the scalar glueballs and realized that, first,
the isothermality is automatically achieved, and, second, that the predicted values are close to the unquenched lattice estimations
of $T_c$. The accordance of the unquenched glueball spectrum fit giving $T_c\simeq 175 \pm 15$~MeV appears particularly
successful in our view.

In conclusion, we suppose that our analysis contains several new observations concerning the 
calculation of the deconfinement temperature in AdS/QCD.
At the same time, it is not meant to be a criticism of \cite{Herzog:2006} and its followers (as we exploited a similar treatment in \cite{estim_deconf}). Rather, we consider this as a possibility to 
clarify the methodology and broaden the horizons of (semi-)quantitative bottom-up holographic estimations.

We do not provide any new theoretical insight on why the first order Hawking--Page phase transition should reproduce the crossover nature of the QCD deconfinement
process or on the string theory configurations leading to the GSW models. But we attempted, within some of these models, to make an honest estimation
of the quantity of much discussion in the modern physics in these models that may not be perfect but do retain some predictive 
power.

For the development of this work we see several directions. First, we may try to be not limited
by only scalar glueballs and consider pomeron/odderon Regge trajectories. Second, we may study various ways to include
other axes (vector, axial and isospin quark chemical potentials, magnetic field)
to the holographically generated phase diagram.

\bibliography{glueball_temperature}
	
\end{document}